\documentclass[aps,prl,twocolumn,showpacs,superscriptaddress,showpacs]{revtex4}

\usepackage{graphicx}
\usepackage{color}

\begin{document}
\title{Simultaneous quantization of bulk conduction and valence states through adsorption of nonmagnetic impurities on  Bi$_2$Se$_3$}
\author{Marco Bianchi}
\affiliation{Department of Physics and Astronomy, Interdisciplinary Nanoscience Center, Aarhus University, 8000 Aarhus C, Denmark}
\author{Richard C. Hatch}
\affiliation{Department of Physics and Astronomy, Interdisciplinary Nanoscience Center, Aarhus University, 8000 Aarhus C, Denmark}
\author{Jianli Mi}
\affiliation{Center for Materials Crystallography, Department of Chemistry, Interdisciplinary Nanoscience Center, Aarhus University, 8000 Aarhus C, Denmark}
\author{Bo Brummerstedt Iversen}
\affiliation{Center for Materials Crystallography, Department of Chemistry, Interdisciplinary Nanoscience Center, Aarhus University, 8000 Aarhus C, Denmark}
\author{Philip Hofmann}
\affiliation{Department of Physics and Astronomy, Interdisciplinary Nanoscience Center, Aarhus University, 8000 Aarhus C, Denmark}
\email[]{philip@phys.au.dk}


\date{\today}
\begin{abstract}
Exposing the (111) surface of the topological insulator Bi$_2$Se$_3$ to carbon monoxide results in strong shifts of the features observed in angle-resolved photoemission. The  behavior is very similar to an often reported ``aging'' effect of the surface and it is concluded that this aging is most likely due to the adsorption of rest gas molecules. The spectral changes are also similar to those recently reported in connection with the adsorption of the magnetic adatom Fe. All spectral changes can be explained by a simultaneous confinement of the conduction band and valence band states. This is only possible because of the unusual bulk electronic structure of Bi$_2$Se$_3$. The valence band quantization leads to spectral features which resemble those of a band gap opening at the Dirac point. 
\end{abstract}

\maketitle

The existence of stable, topologically protected metallic states on the surfaces of certain bulk insulators, combined with the many special properties of these surface states, is currently generating considerable interest in condensed matter physics \cite{Zhang:2008,Hasan:2010}. One remarkable property is the spin texture of the surface states which prevents back-scattering \cite{Pascual:2004,Roushan:2009} such that the surface state electrons do not suffer localization by weak disorder \cite{Fu:2007c,Qi:2008} and one-dimensional edge states do not undergo any scattering \cite{Konig:2007}. However, these restrictions only hold as long as time-reversal symmetry is preserved, such that the introduction of a magnetic field or magnetic impurities should strongly alter the situation.

Angle-resolved photoemission spectroscopy (ARPES) is an ideal tool to study the metallic states on topological insulators as it is surface-sensitive and can give detailed information about the spectral function. 
ARPES experiments have, for instance, shown that topological insulators indeed support surface states with a Dirac-cone like dispersion and a Fermi surface topology consistent with theoretical predictions \cite{Hsieh:2008,Hsieh:2009b}. Adding spin-resolution has verified that the surface states have the expected spin texture  \cite{Hsieh:2009,Hsieh:2009c}. The first material experimentally shown to be a topological insulator in this way was a disordered Bi-Sb alloy but more recently most experiments have been performed on the prototypical materials Bi$_2$Se$_3$ and Bi$_2$Te$_3$ which have a simpler electronic structure (a single Dirac cone) and a substantially larger bulk band gap \cite{Zhang:2009,Xia:2009,Hsieh:2009b}. 

There are a number of important unresolved issues linked to the basic electronic structure of these materials. The first is an `aging' effect which typically takes place over several hours after obtaining a fresh surface by cleaving the crystals \cite{Hsieh:2009c}. The aging manifests itself as an increasing downward bending of the bands and the eventual appearance of quantized two-dimensional states in the conduction band \cite{Bianchi:2010b}. The origin of the effect is unclear but a surface relaxation has been discussed as a possible reason \cite{Hsieh:2009b,Hsieh:2009c}. 

Another open question is the effect of surface or bulk impurities on the surface electronic structure. 
Especially interesting are magnetic impurities since these are expected to break time-reversal symmetry, thus opening channels for back scattering
and localization of the electrons in the topological state.
They can also give rise to a distinct spectroscopic signature which is an opening of a gap at the Dirac point (DP) of the topological state, i. e. the transition from massless to massive fermions. Such gap-openings have indeed been observed for both bulk and surface impurities \cite{Chen:2010,Wray:2011}. For nonmagnetic impurities, on the other hand, one would naively expect a mere doping effect and an increase of defect scattering but distinct spectral changes in the vicinity of the DP  have also been predicted recently \cite{Biswas:2010}.

In this paper, we show that these two issues are closely related. The adsorption of carbon monoxide, a nonmagnetic molecule commonly present in the rest gas of a vacuum recipient, can lead to a strong modification of the surface electronic structure. It brings about the aforementioned aging effect by inducing a strong band bending. Combined with the unusual valence electronic structure of Bi$_2$Se$_3$, this band bending can become strong enough to induce a \emph{simultaneous} quantization of conduction band (CB) and valence band (VB) states, an effect not known from  any other semiconductor surface. The quantization of the valence states in the vicinity of the DP can resemble a gap-opening and the spectral appearance of the CO-induced changes is very similar to that brought about by Fe adsorption \cite{Wray:2011}. 

ARPES experiments were performed on $in$ $situ$-cleaved single crystals of Bi$_2$Se$_3$ at the ASTRID synchrotron radiation facility. The bulk crystals were doped with calcium in order to bring the bulk Fermi level into the gap, very close to the surface state DP \cite{Hor:2009}. Photoemission spectra were measured with an angular resolution of 0.13$^{\circ}$ and a combined energy resolution better than 15~meV. Spectra were taken along the $\bar{K}\bar{\Gamma}\bar{K}$ azimuthal direction. The sample temperature was 65~K. Carbon monoxide was dosed while taking photoemission spectra with a partial pressure of $8 \times 10^{-9} $~mbar. The data shown for the CO-uptake and the photon energy scan are merely a subset from much more extensive data sets. These complete data sets are provided as movies in the supplementary online material \cite{SOM}. 

Figure \ref{fig:1} shows the evolution of the surface electronic structure as it is exposed to CO. We first discuss the changes near the Fermi level and in the CB. Immediately after cleaving the sample the DP is very close to the Fermi level at a binding energy of 46(5)~meV. This uptake experiment was started about an hour after the cleave and the DP had moved to a binding energy of 102(5)~meV (see Fig. \ref{fig:1}(a)). Still, only the topological state is present at the Fermi surface. As the CO coverage increases, the DP moves rapidly to higher binding energies, indicative of an increasing band bending near the surface. Further states emerge at the Fermi level and move to higher binding energies (see Fig. \ref{fig:1}(b),(c)). These states have been shown to be quantized sub-bands of the CB \cite{Bianchi:2010b}. For a sufficiently strong potential gradient, the quantum well states show a considerable Rashba splitting \cite{King:2011}, similar to that observed for the surface states of Au(111) \cite{Lashell:1996}.

The VB region is also affected by the CO adsorption. Immediately after cleaving, a single  M-shaped state is  observable in this region but after a short time two such states can be identified, as in Fig. \ref{fig:1}(a). The state with the higher binding energy has been assigned to a surface state as it shows no dispersion with $k_z$ \cite{Bianchi:2010b}. The state with the lower binding energy and the similar, well-separated features which appear for a higher CO coverage, all fall in the energy region of the projected valence band \cite{Eremeev:2010b}. We will later argue that these are quantum well states (QWS) formed in the valence band and that this is also a possible alternative explanation for the state with the highest binding energy. Fig. \ref{fig:2} shows a high-coverage data set over a wider energy range and with better statistics. The QWS are considerably sharper than the topological state such that it is difficult to determine the exact position of the DP with respect to the QWS. Extrapolating the position of the DP from the dispersion at lower binding energy suggests that it lies in between the QWS, consistent with coverage-dependent position of the states in Fig. \ref{fig:1}(c). 

\begin{figure}
\begin{center}
\includegraphics[width=\columnwidth]{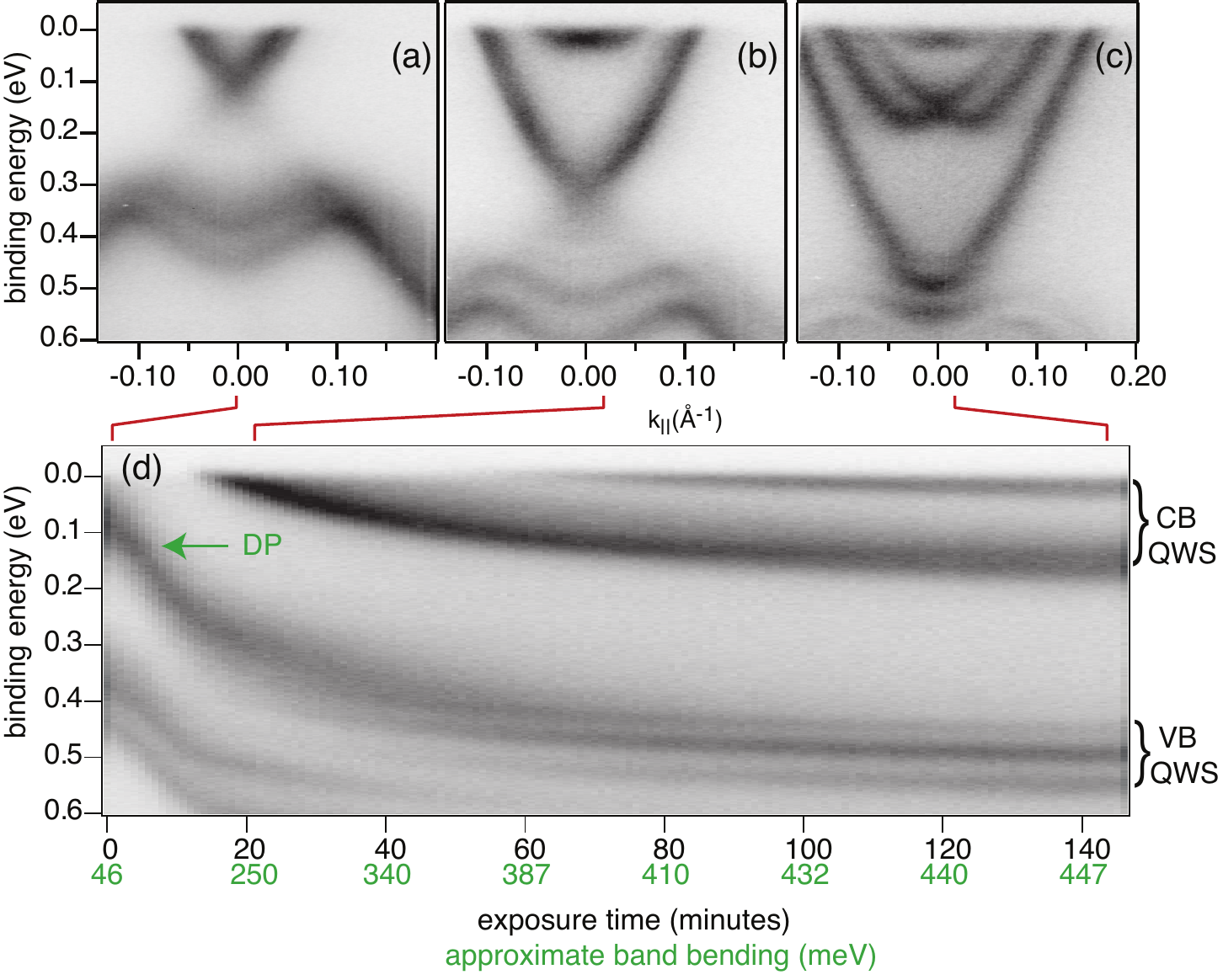}
\caption{(color online). ARPES spectra taken during the exposure of the surface to carbon monoxide ($h\nu=16$~eV). (a) - (c) photoemission intensity for different exposures as a function of binding energy and k-vector parallel to the surface (dark corresponds to high intensity). (d) cut through the center of the entire series of such images, illustrating the development of the different states with exposure time. The states marked in (d) are the conduction band and valence band quantum well states (CB QWS and VB QWS) as well as the Dirac point of the topological surface state (DP)  \cite{SOM}. } 
\label{fig:1}
\end{center}
\end{figure}

\begin{figure}
\begin{center}
\includegraphics[width= 0.7 \columnwidth]{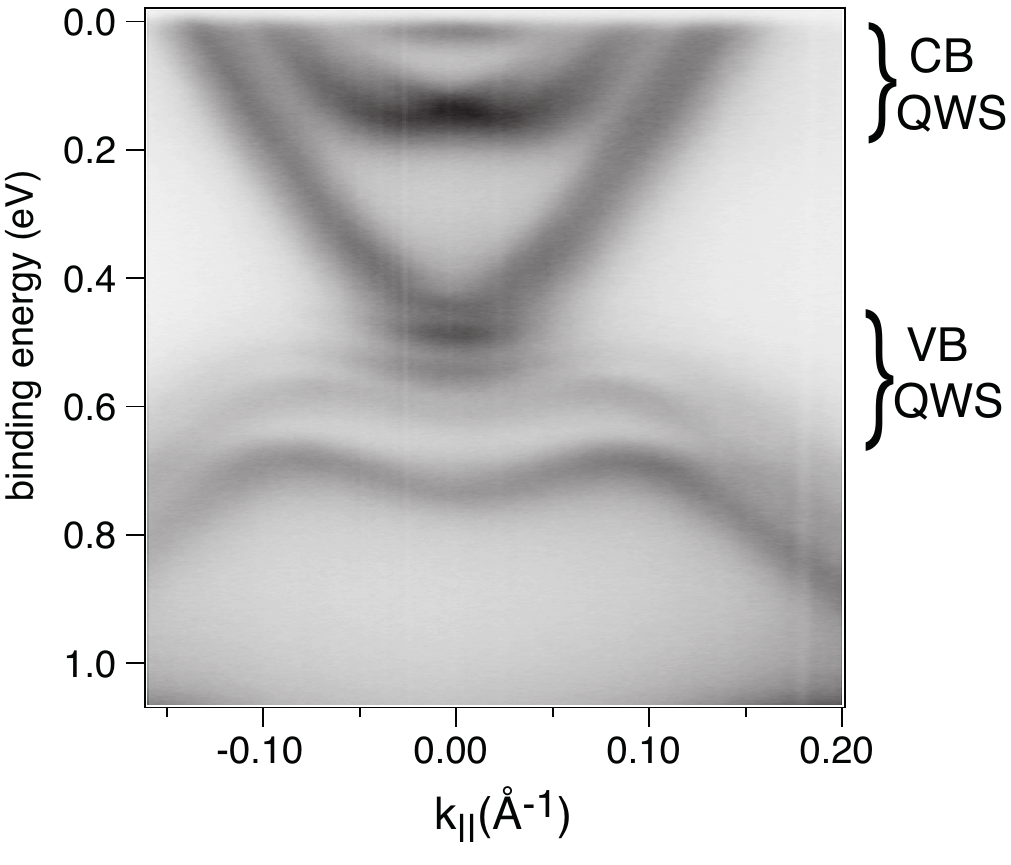}
\caption{Detailed ARPES spectrum of the valence band quantum well states for the full CO exposure reached in the end of the uptake in Fig. \ref{fig:1}  ($h\nu=16$~eV). } 
\label{fig:2}
\end{center}
\end{figure}

In the following, we argue that the M-shaped states in the VB arise from a quantization. Such a quantization is counter intuitive: while a downward band bending near the surface is expected to confine the CB states, an upward bending would be needed to confine the VB states \cite{Takeda:2005,Kim:2010} and thus a simultaneous confinement of both bands does not appear to be possible. In the particular case of Bi$_2$Se$_3$, however, this simple picture does not hold because the top part of the VB exists only in a narrow energy region near the surface Brillouin zone center \cite{Xia:2009,Eremeev:2010b}. 

\begin{figure}
\begin{center}
\includegraphics[width= \columnwidth]{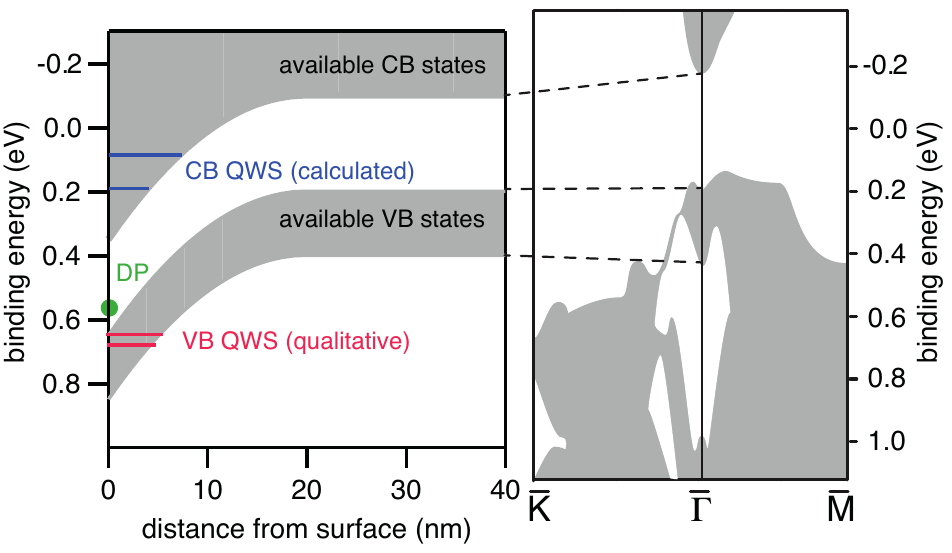}
\caption{(color online). Semi-quantitative model for the quantum confinement of the CB and VB states. The band bending corresponds to the value for the full coverage in Fig. \ref{fig:1}. The width of the space charge layer has been estimated by requiring the presence of two occupied quantum well states in the CB with similar binding energies as observed experimentally and integrating the Schr\"odinger equation numerically. The green marker denotes the position of the Dirac point. The position of the quantum-confined states in the VB is merely a qualitative sketch. Confining these states becomes possible because of the narrow total width of the VB near the surface Brillouin zone center. This is illustrated on the right part of the figure which gives the calculated bulk band structure projection adapted from Ref. \cite{Eremeev:2010b}. } 
\label{fig:3}
\end{center}
\end{figure}

We determine the approximate electrostatic potential near the surface from the position of the quantum well states in the CB. To this end, we use a simple  Shottky model of the space charge zone. A constant charge density is assumed to be present between the surface ($z=0$) and a certain depth ($z=\Delta$). From this, the electrostatic potential is calculated via the Poisson equation. The model has two parameters, the total band bending and the width of the charge layer $\Delta$. 
The total band bending can be inferred from the position of the topological state at off-normal emission where the state can be clearly observed even for high coverages. It is found to be $\approx$ 450~meV. $\Delta$ is an adjustable parameter.
We then numerically solve the  Schr\"odinger equation in this potential and  adjust $\Delta$ such that the number of the occupied solutions is the same as the number of observed QWS in the CB (two) and their energy eigenvalues are in approximate agreement with the experimental values which are 334(5)~meV and 472(5)~meV above DP. The resulting potential and the calculated solutions are given in Fig. \ref{fig:3}.

Note that no special topological significance can be assigned to the quantum states in the CB, not matter if they are Rashba-split
 \cite{King:2011} or not. First of all, the QWS character of these states means that they are simply part of the CB and therefore necessarily  ``topologically trivial''. Moreover, the topological arguments do not hold in the energy region of projected bulk states but merely in the gaps between them.

In the absence of effects like near-surface band shrinkage \cite{King:2010}, the VB bending merely follows that of the CB. 
It can be seen that the magnitude of this band bending is more than twice as large as the calculated total width of the VB (around 200~meV \cite{Xia:2009,Eremeev:2010b}) and this makes it possible to confine VB-derived states as indicated in the figure. Note, however, that it is not possible to calculate the energies of these quantum confined states in a simple manner similar to the CB because the effective mass of these states changes strongly with energy and distance to the surface. 

\begin{figure}
\begin{center}
\includegraphics[width=\columnwidth]{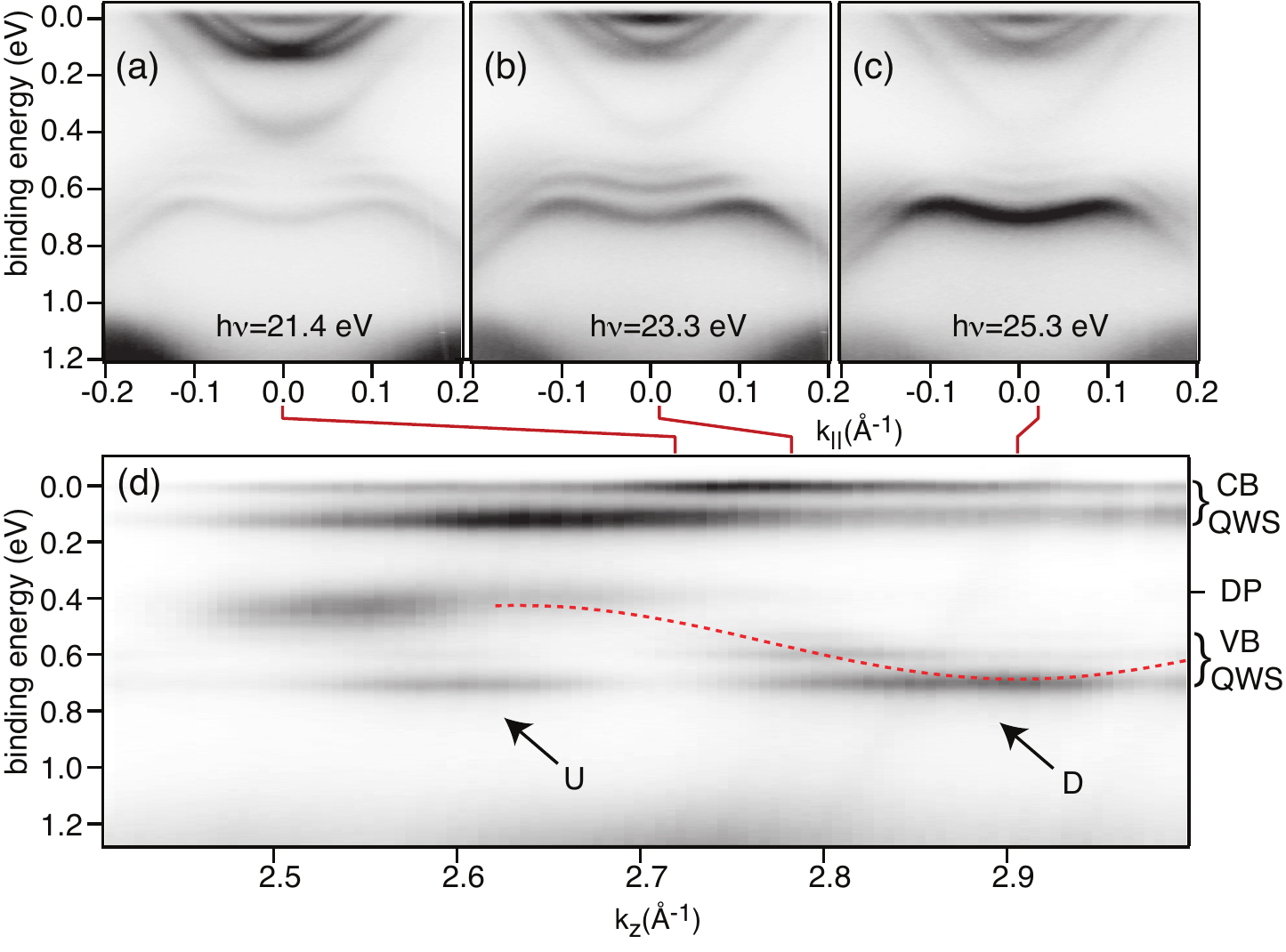}
\caption{(color online). Photon energy scan to probe the $k_z$ dependence of the observed states.  (a) - (c) photoemission intensity for selected photon energies  \cite{SOM}. (d) cut through the center of the entire series of such images, illustrating the dispersion of the states at normal emission. Neither the CB nor the VB QWS show any dispersion but the VB QWS closely track the dispersion of the VB which is observed in the absence of quantization, here given as a dashed line to guide the eye \cite{Bianchi:2010b}. The branch marked D corresponds to the direct photoemission while the branch marked U involves an umklapp process with a surface reciprocal lattice vector. } 
\label{fig:4}
\end{center}
\end{figure}

There are several experimental results underpinning the above interpretation of the M-shaped states in the VB. The first is the absence of dispersion with $k_{z}$ which is indicative of a state's two-dimensional nature. This can be directly accessed via a photon energy scan. We chose an energy range between 14 and 28.3~eV in which the dispersion of the VB is most clearly identified. The result of such a photon energy scan, taken for a slightly lower CO coverage than the maximum in Fig. \ref{fig:1} is given in Fig. \ref{fig:4}. The $k_z$ axis in (d) has been determined using free electron final states. Details are given in Ref. \cite{Bianchi:2010b} and Fig.~2 in this reference shows the corresponding scan for a smaller band bending and non-confined VB states. The non-dispersive character of the QWS is evident, but it is also interesting to note the energies at which the emission intensity from these states is resonantly enhanced. This happens at the same energies as the emission from the corresponding non-confined states, either through direct photoemission (marked by a D in the figure) or via a surface umklapp process (marked by an U), clearly illustrating the origin of the QWS as being derived from the VB states with the corresponding energies.

The idea of VB confinement is further supported by the following consideration. Following the model laid out in Fig. \ref{fig:3}, a necessary condition for the appearance of quantized VB sub-levels is that the band bending is larger than the total width of the VB. The appearance of the quantized valence band features in Fig. \ref{fig:1}(d) is indeed consistent with this, as they are only  observed for a band bending larger than 250~meV or so. Note, however, that the states at the bottom of the valence band could already be confined at a smaller band bending. This is  particularly true for the M-shaped state which is observed before CO adsorption and which has previously been interpreted as a surface state \cite{Bianchi:2010b}. Given its proximity to the calculated band edge, an alternative interpretation as a QWS is equally reasonable. 

It is very interesting to compare our results to those obtained by the adsorption of Fe on the surface of Ca-doped Bi$_2$Se$_3$ \cite{Wray:2011}. The similarities are striking. In the CB very similar Rashba-split quantum well states are observed which are interpreted as additional surface states. The VB shows a series of M-shaped features, very similar to what is reported here. The states are, however, not interpreted in the same way but rather in terms of a gap-opening at the DP. In view of the present results it appears reasonable to argue that there is very little difference between nonmagnetic and magnetic impurities in these two cases and both merely lead to doping and strong band bending. Note, however, that adsorbed Fe atoms are not necessarily magnetic in the first place. It would have to be confirmed that the Fe adatoms retain a significant magnetic moment after adsorption and a band gap opening at the DP would only be expected for an ordering of the moments perpendicular to the the surface \cite{Abanin:2011}.

The overall good agreement with the presented picture of QWS formation does not rule out other interpretations for the spectral changes near the DP. As mentioned above, the precise position of the DP in the region of the VB QWS is not clear. It seems to be placed inside the regions of QWS whereas it appears to be just outside the valence band region for the pristine surface \cite{Xia:2009,Bianchi:2010b}. This could have several reasons. One is that the DP is also degenerate with the VB top for the pristine surface and that this is merely hard to observe. 
Confining the VB states into QWS may simply facilitate the observation of their precise energies. It is also possible that the DP shifts to higher energies than the VB because it is more localized in the immediate vicinity of the surface and thus placed in a stronger average field.  A more interesting alternative interpretation would be that the large quantity of CO gives rise to spectral changes close to the DP as predicted by Biswas and Balatsky \cite{Biswas:2010} who show that sharp resonances in the density of states are possible in the vicinity of charged (but nonmagnetic) impurities. 

Summarizing, our results provide an interpretation of the multiple M-shaped features in the VB region of Bi$_2$Se$_3$ as multiple QWS. The fact such quantized states can exist in the case of a downward band bending is highly unusual and, to the best of our knowledge, a unique property of this, and probably similar, materials. The observed spectral changes, including an apparent `gap opening' at the DP,  are very similar to those reported in the case of Fe adsorption on  Bi$_2$Se$_3$ \cite{Wray:2011}, raising the question if they are indeed due to the magnetic character of the impurity in the latter case. Finally, the fact that CO is common in the rest gas of a vacuum recipient and that it induces the strong band bending reported here suggest that the observed aging of several topological insulator surfaces is caused by adsorption and could possibly be avoided  by a thin protective layer on the surface.

We acknowledge stimulating discussions with Ph.~D.~C.~King, A.~A.~Khajetoorians, J.~Wiebe, S.~V.~Eremeev and E.~V.~Chulkov as well as financial support by the Lundbeck foundation and the Danish National Research Foundation.



\end{document}